\documentclass[twoside,slac_one]{revtex4}
\usepackage{graphicx}
\usepackage{fancyhdr}
\usepackage{amsmath} 
\usepackage{bm}
\usepackage{amsxtra}
\usepackage{amssymb}
\usepackage{amsthm}
\usepackage{latexsym}
\usepackage{lscape}

\def\BR{{\cal B}}

\def\Hpm{ {H}^{\pm\pm}}

\def\HpmL{ {H}^{\pm\pm}_{L}}
\def\HpmR{ {H}^{\pm\pm}_{R}}

\def\MET{{p\kern -0.4em/_{T}}}

\begin{document}

\title{Search for charged and doubly-charged Higgs boson production in proton-antiproton collisions at   $\sqrt{s}$ =  1.96 TeV}

%

\author{L. Suter}
\affiliation{School of Physics and Astronomy, University of Manchester, Manchester, UK}

\begin{abstract}
We present searches for charged Higgs production in decays of top quarks and also pair production of doubly charged Higgs boson decaying to di-tau, di-muon, and muon $+$ tau final states. The searches are performed in proton-antiproton collisions at $\sqrt{s}$ = 1.96 TeV using an integrated luminosity of up to 7 fb$^{-1}$ collected by the CDF and D0 experiments at the Fermilab Tevatron Collider. We find no evidence for charged Higgs production and set limits on the production cross-section for a variety of theoretical models. This represents the first search for pair production of doubly-charged Higgs bosons decaying into tau leptons at a hadron collider.
\end{abstract}

\maketitle

\thispagestyle{fancy}


\section{Introduction}

Charged and doubly charged Higgs appear in various extensions to the Standard Model, SM. In order to generate a charged Higgs one needs to have two Higgs doublets, describing the Higgs sector, compared to the one in the SM. One set of theories of particular interest are supersymmetric, SUSY, theories. These are extensions to the SM which predicts a new symmetry between bosons and fermions. This theory has several advantages over the SM such as the introduction of a dark matter candidate, a solution to the hierarchy problem and the potential for GUT scale unification. In its simplest form it is the MSSM, the minimal supersymmetric standard model, which after electroweak symmetry breaking predicts 5 Higgs boson, 3 neutral and 2 charged. Charged Higgs also appear in extended SUSY models such as the NMSSM, the next to minimal supersymmetric standard model ~\cite{nmssm}. At tree level the Higgs sector of the MSSM can be described by two parameters which are chosen to be, tan$\beta$, the ratio of the vacuum expectation of the two Higgs doublets and, $M_A$, the mass of the pseudo-scalar Higgs boson.  Beyond tree level radiative corrections bring in dependance on more than just $M_A$ and tan$\beta$ so limits are set at specific benchmark regimes in the $M_A$-tan$\beta$ plane.  
To produce a double charged Higgs the SM must be extended further with the inclusion of a Higgs triplet, this occurs in models such as the Left Right Symmetry Model  ~\cite{lrsym} and the Little Higgs models ~\cite{littlehiggs}.

\section{Charged Higgs}

The decays of charged Higgs depends on both the model considered and the mass range predicted. Various scenarios have been studied at both D0 and CDF, for non-extended two Higgs doublet models, such as the MSSM, the charged Higgs is expected to couple strongest to the top quark. Therefore analysis studied can be separated in cases where the charged Higgs is lighter than the top quark and where it is more massive. 

\subsection{Two Higgs doublet models: $M_{top} > M_{H}$}

For the case where the mass of the charged Higgs is smaller than the mass of the top quark, the production of a charged Higgs at the Tevatron will be via $ t \rightarrow H^+\bar{b}$. The charged Higgs can either decay leptonically, hadronically or to a mixed hadron lepton final state. 
This additional decay of the top quark will compete with the SM process $t \rightarrow W^+\bar{b}$. Due to the different decays of the $W^+$ and $H^+$ a search for the charged Higgs can be preformed by looking at a difference in the distributions of events between the final states as compared to what is predicted by the standard model. 
Searches had been performed for a leptophobic model where it is assumed that the BR$(H^+ \rightarrow c\bar{s}) \approx 1$.  This was preformed at CDF with 2.2 fb$^{-1}$ of integrated luminosity and at D0 with 1.0 fb$^{-1}$ ~\cite{cdf_1, d0_1}.
D0 has also conducted a search for both the leptophilic case where it is assumed that BR$(H^+ \rightarrow \tau\nu) \approx 1$ and a strangephilic model where BR$(H^+ \rightarrow c\bar{s})$ +  BR$(H^+ \rightarrow \tau\nu) \approx 1$ ~\cite{d0_1} with 1.0 fb$^{-1}$ of integrated luminosity.

\begin{figure}[ht]
\centering
\includegraphics[width=80mm]{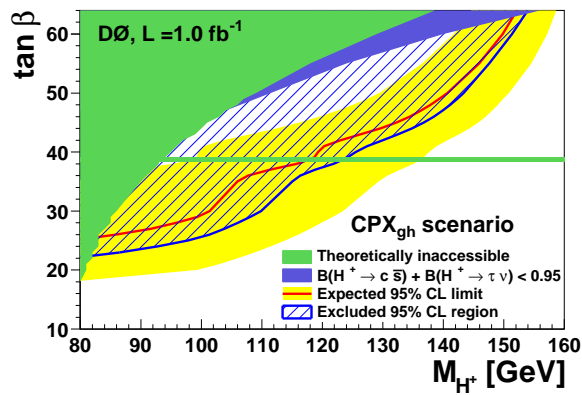}
\caption{95\% confidence limits on the cross section $\times$ branching ratio of a charged MSSM Higgs boson shown in the $M_{H^+}$-tan$\beta$ plane for a strangephilic model of charged Higgs coupling. The expected limit is shown by the red line and the hatched blue region is the region excluded at the 95\% confidence level. } \label{figure_one}
\end{figure}

\subsection{Two Higgs doublet models: $M_{H} > M_{top}$}

If it is assumed that the mass of the charged Higgs is greater than the mass of the top quark, then the charged Higgs will be produced through $p\bar{p}$ annihilation and will decay to a $t\bar{b}$ pair.  This is the same final state as the standard model process $W^+ \rightarrow t\bar{b}$ therefore the rate of single top production at the Tevatron compared to that predicted by the SM would be effected and a bump would be expected in the invariant mass distribution of the $t\bar{b}$ final state. D0 performed a search for a charged Higgs boson is this region between the masses of $180 < M_{H^+} < 300$ GeV using 0.9 fb$^{-1}$ of integrated luminosity. No evidence of a charged Higgs boson was found and limits where set on $\sigma(p\bar{p} \rightarrow H^{+}) \times$ BR$(H^{+} \rightarrow t\bar{b})$ for three different types of two Higgs doublet model.  In Type One two Higgs doublet models only one of these doublets couples to fermions and a region in the plane of the charged Higgs mass and tan$\beta$ and be excluded ~\cite{d0_2}. This can be seen in Figure 2.  

\begin{figure}[ht]
\centering
\includegraphics[width=80mm]{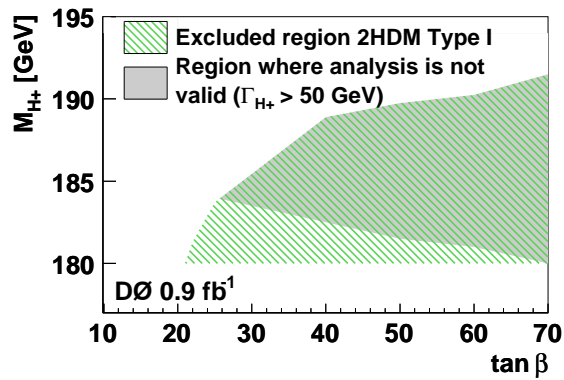}
\caption{95\% confidence limits on the cross section $\times$ branching ratio of a charged MSSM Higgs boson shown in the $M_{H^+}$-tan$\beta$ plane for a heavy charged Higgs decaying into $t\bar{b}$ pairs. The region excluded is shown in green. } \label{figure_two}
\end{figure}

\subsection{Extended two Higgs doublet models}

Charged Higgs boson also appear in extended Higgs models such as the NMSSM, this model extends the MSSM by the addition of a singlet Higgs boson on top on the two doublets in the MSSM. After electroweak symmetry breaking this results in 7 Higgs bosons, 5 neutral and 2 charged. The NMSSM has several advantages over the MSSM, it provides a natural scale for the $\mu$ parameter,  supersymmetric mass parameter, it alleviates the little hierarchy problem and for regions of the NMSSM parameter space where the mass the lightest pseudo-scalar Higgs boson, $a$,  is less than twice the mass of the $b$ quark, $m_a < 2m_b$ then the LEP limits can be avoided ~\cite{nmssm}. Within the NMSSM,  for large regions of the parameter space Higgs-to-Higgs decays dominate over decays to other particles. 
A search has been performed at CDF for a NMSSM charged Higgs boson with 2.7 fb$^{-1}$ of integrated luminosity ~\cite{nmssm_cdf}. It was assumed that the charged Higgs will be produced from the decay of a top quark and will decay into a light psuedo-scalar Higgs boson.  The mass of  scalar Higgs is assumed to be less than twice the mass of the $b$ quark and hence it will to decay to a $\tau\tau$ pair, $t\rightarrow H^+ b \rightarrow W^{\pm}ab \rightarrow W^{\pm} \tau\tau b$. Limits are determined on the $t \rightarrow H^+ b$ branching ratio for various $H^+$ and $a$ masses by preforming a fit to the isolated track $p_T$ spectrum.  The determined limits on the BR($t \rightarrow H^{\pm} b)$ as a function of the charged Higgs mass are shown in Figure 3. 

\begin{figure}[ht]
\centering
\includegraphics[width=80mm]{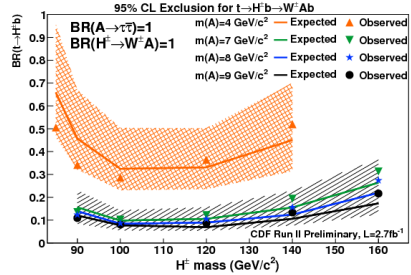}
\caption{ 95\% confidence limits on the BR($t \rightarrow H^{\pm} b)$ as a function of the charged Higgs mass. Four different values of the pseudo-scalar Higgs mass are shown. } \label{figure_three}
\end{figure}

\section{Doubly charged Higgs}

Doubly charged Higgs appear in extensions to the SM in which introduce a Higgs triplet. After electroweak symmetry breaking this results in Higgs boson with a double charge. Higgs triplet models that have been covered by analyses at the Tevatron included the Left Right  Symmetry model, which predicts a new symmetry between left and right handed particles ~\cite{lrsym}. In this model there is both a left-handed and a right-handed doubly charged Higgs boson, $H^{\pm\pm}_L, H^{\pm\pm}_R$ where is predicted that the production cross section of the right handed Higgs is approximately half that to the left-handed state due to the difference in the coupling to the $Z$ boson where the left- and right-handedness of the doubly charged Higgs boson refers to the handedness of the particles that couples to and not of the Higgs itself. 
Higgs triplets also appear in Little Higgs models, with a type 2 see-saw mechanism where they are used as a production mechanism for neutrino masses~\cite{littlehiggs}. For a normal hierarchy of neutrino masses below approximately 10 meV, these models predict that the doubly charged Higgs will have approximately equal branching ratios to $\mu\mu, \tau\tau$ and $\mu\tau$ final states ~\cite{equalBR}. In $(3-3-1$) gauge symmetric models which predict additional heavy exotic quarks and leptons that provide anomaly cancellations, double charged Higgs that decays dominantly to taus ~\cite{331}. D0 has performed a search looking for the pair preproduction of $H^{\pm\pm}$ decaying into $\tau^{\pm}\tau^{\pm}, \mu^{\pm}\mu^{\pm}$ and $\mu^{\pm}\tau^{\pm}$ final states using up to 7.0 fb$^{-1}$ of integrated luminosity. Detailed description of this analysis can be found in ~\cite{me}. This is the first search for doubly charged Higgs decays to a pure hadronic tau final state, $H^{\pm\pm} \rightarrow \tau^{\pm}\tau^{\pm}$,  at a hadron collider.  Limits were set for both model dependent scenarios. i.e.  assuming a equal BR to $\mu\mu, \tau\tau$ and $\mu\tau$ final states as predicted in ~\cite{equalBR} and for model independent ones. These were BR$(H^{\pm\pm} \rightarrow \tau^{\pm}\tau^{\pm}) =1$, BR$(H^{\pm\pm} \rightarrow \mu^{\pm}\mu^{\pm}) =1$, BR$(H^{\pm\pm} \rightarrow \mu^{\pm}\tau^{\pm}) =1$ and BR$(H^{\pm\pm} \rightarrow \tau^{\pm}\tau^{\pm})$ + BR$(H^{\pm\pm} \rightarrow \mu^{\pm}\mu^{\pm}) =1$, where BR$(H^{\pm\pm} \rightarrow \tau^{\pm}\tau^{\pm})$ is ranged over 10 different value from 100\% BR to taus to 100\% BR to muons. All the model independent limits are determined for both and right- and left-handed Higgs. 

\subsection{Analysis Summary} 

This analysis starts with the identification of muons and hadronic taus are identified. For a reconstructed muon, hits in the muon chambers are matched to tracks in the central tracking chambers, these are then required to be isolated in both the calorimeters and the central tracking detectors. 
Hadronic tau decays are first split into three types based on the energy clusters in the calorimeters and the number of tracks. Type 1 and 2 are 1-prong decays with energy deposited in the hadronic calorimeter (type 1) or in both the electromagnetic and hadronic calorimeters (type 2). Type 3 taus are 3-prong decays with an invariant mass below 1.7 GeV and energy deposits in the calorimeters. For each of the different types a Neural Network, NN,  is trained using $Z \rightarrow \tau\tau$ decays as signal and multijet events predicted from data as background, to separate hadronic tau decays from jets.  

Events are required to have two hadronic taus, $\tau_{h}$, and at least one isolated muon. The following selection criteria must be satisfied;   the pseudo-rapidity of selected muons and taus must be $|\eta^{\mu}| < $1.6, $|\eta^{\tau_{1,2}}| < 1.5$,  additional taus must have $|\eta^{\tau}| < 2$.  The transverse momentum of the objects must satify $p_{T}^{\mu} > 15$ GeV and $p_{T}^{\tau} > 12.5$. All $\tau_{h}$ and muons are required to be separated by $\Delta R > 0.5$, where  $\Delta R = \sqrt{ (\Delta\phi)^2 + (\Delta\eta)^2}$, for the two leading $p_T$ taus then $\Delta R > 0.7$. The sum of the charges of the highest $p_T$ muon and  the two highest $p_T$ taus is required to be $Q = \sum_{i=\mu, \tau_1, \tau_2}  q_i = \pm 1$, as is expected for signal. 
After application of the described cuts, the main backgrounds contributions are from diboson and $Z \rightarrow \tau\tau$ processes, with smaller contributions arising from $Z \rightarrow \mu\mu$, $W$+jets, $t\bar{t}$ and $Z \rightarrow ee$.  All background process are simulated with ALPGEN ~\cite{alpgen} with showering and hadronization provided by PYTHIA ~\cite{pythia} except for diboson and signal events which are generated by PYTHIA.  Tau decays are simulated using TAUOLA which correctly model tau polarization ~\cite{tauola}.  For all MC samples GEANT ~\cite{geant} is used to correct for detector effects.  The contribution due to multijet background is determined to be negligible.

To enhance the discrimination of the signal sample over the background the selected events are split into four different samples. These are based on the number of tau and muons in the final state and the charged correlation between the the two taus.  For two samples it is required that there is exactly two taus and one muon and these are split further depending on whether the charges of the two taus are the same or opposite. The other two samples require exactly 3 taus and 1 muon or exactly 2 taus and 2 muons.  The invariant mass of the two highest $p_T$ candidates for the two samples with exactly 2 taus and 1 muon are shown in Figure 4. 

\begin{figure*}[ht]
\centering
\includegraphics[width=80mm]{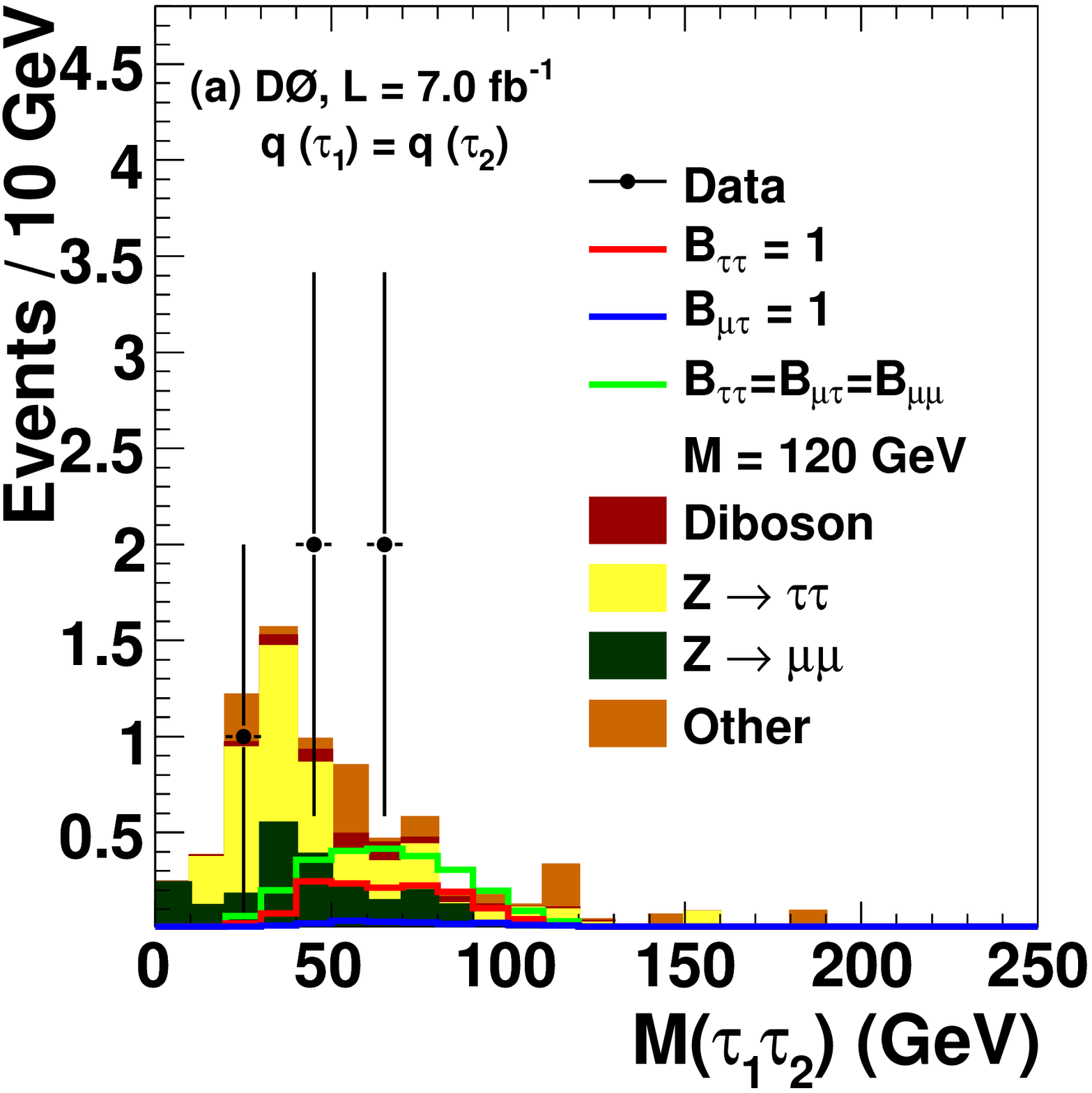}
\includegraphics[width=80mm]{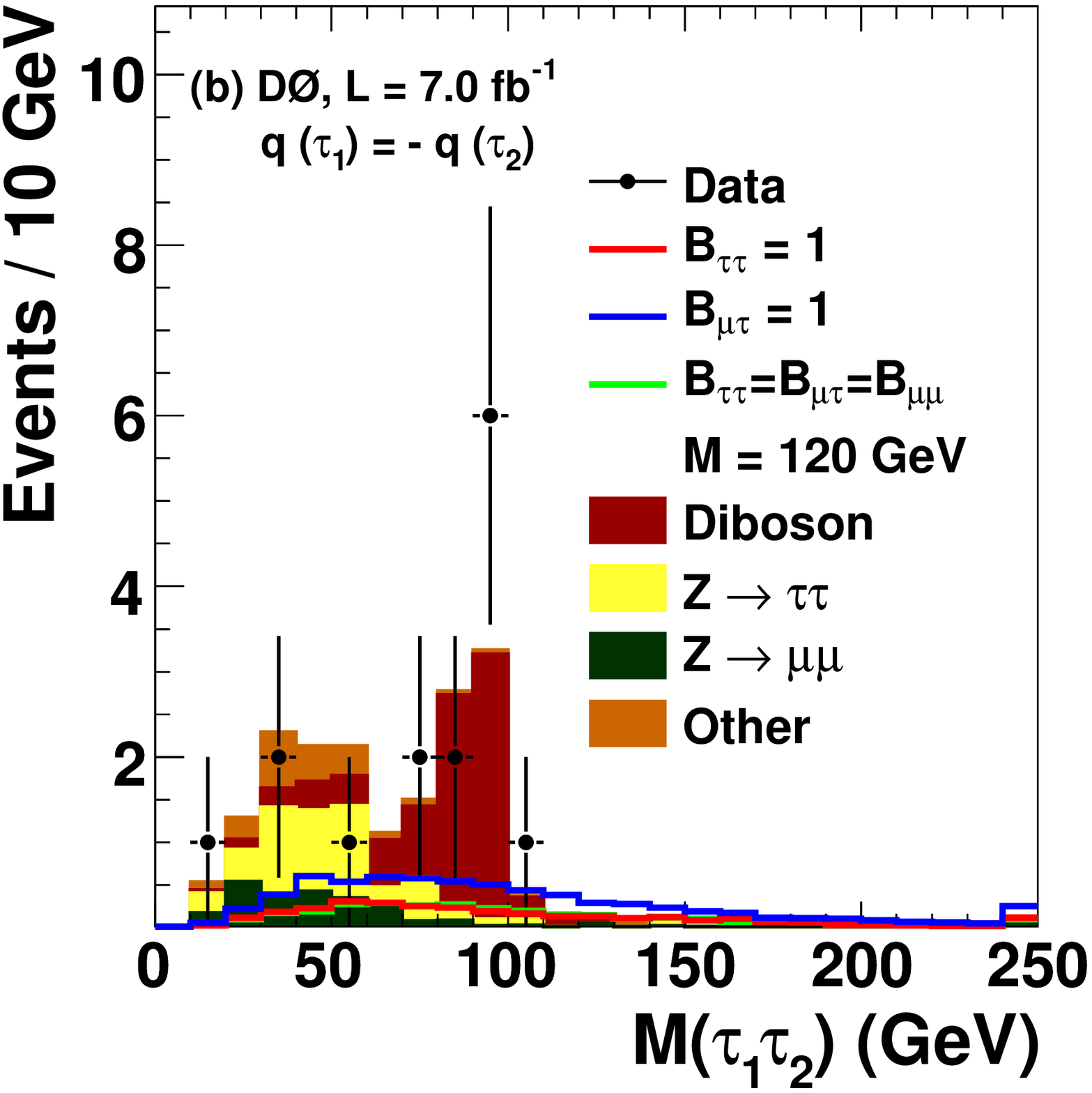}
\caption{The invariant mass of the two highest $p_T$ taus for the selection with 2 taus and 1 muon for (a)  $q(\tau_1) = q(\tau_2)$ (b) $q_(\tau_1) = - q(\tau_2)$. The crosses show the data, the stacked histogram the simulated background and the solid lines expected signal for three values of the $H^{++}$ branching ratio and a mass of 120 GeV. } \label{figure_four}
\end{figure*}

The four different samples are dominated by different background processes hence splitting improves the sensitivity to the signal. The like charged sample is dominated by $Z/\gamma^*$ events, mainly $Z/\gamma^* \rightarrow \tau\tau$ with contributions from $Z/\gamma^* \rightarrow \mu\mu$ and $W$+jets. The opposite charged sample is dominated by diboson events mainly $WZ \rightarrow \mu\nu e^+ e^-$ with electrons misidentified as taus. Smaller contributions arise from $Z/\gamma^*$ and $W$+jets events. Table I shows the expected number of signal and background events for the four different signal samples.

\begin{table*}[t*]
\caption{Numbers of events in data, predicted background, and expected 
signal for $M(\HpmL)=120$~GeV, 
assuming the NLO calculation of the signal cross section for
 BR$(\HpmL\to\tau^{\pm}\tau^{\pm})=1$,
 BR$(\HpmL\to\mu^{\pm}\tau^{\pm})=1$,  and
BR$(\HpmL\to\tau^{\pm}\tau^{\pm})=$ BR$(\HpmL\to\mu^{\pm}\mu^{\pm})
   = $ BR$(\HpmL\to\mu^{\pm}\tau^{\pm})=1/3$.
 The numbers are shown for the four samples separately, together
 with their total uncertainties. }
\begin{center}
   \begin{tabular}{c|ccccc}
   \hline\hline
& All &  \multicolumn{2}{c}{$N_{\mu}=1$} &
     $N_{\mu}=1$  & $N_{\mu}=2$ \\
 &  &  \multicolumn{2}{c}{$N_{\tau}=2$} 
 &  $N_{\tau}= 3$  & $N_{\tau}=2$ \\
 &  & $q_{\tau_1}=q_{\tau_2}$   &  $q_{\tau_1}=-q_{\tau_2}$ &  &  \\
       \hline
  Signal &&&&&\\
$\tau^{\pm}\tau^{\pm}$&  $6.6 \pm 0.9$  & $1.4 \pm 0.2$ & $3.1 \pm 0.4$  & $1.6\pm 0.2$ & $0.4\pm 0.1$  \\ 
$\mu^{\pm}\tau^{\pm}$	& $13.9 \pm 1.9\phantom{0}$ & $0.3 \pm 0.1$ & $6.8 \pm 0.9$  & $0.4\pm 0.1$  & $6.3\pm 0.9$  \\
Equal BR & $9.5 \pm 1.3$ & $2.5 \pm 0.3$ & $3.1 \pm  1.0$ & $1.2\pm 0.2$  & $2.6\pm 0.4$  \\
    \hline
           Background &&&&&\\
$Z\to\tau^+\tau^-$  & $8.2 \pm 1.1$	& $3.4 \pm 0.5$  & $4.8 \pm 0.7$ & $<0.1$ & $<0.1$  \\
$Z\to \mu^+\mu^-$ & $5.1 \pm 0.7$	& $2.2 \pm 0.3$  & $2.5 \pm 0.4$ & $0.1 \pm 0.1$ & $0.2 \pm   0.1$  \\
$Z\to e^+e^-$         & $0.3 \pm 0.1$	&  $<0.1$  	  & $0.3 \pm 0.1$ & $<0.1$              & $<0.1$   \\ 
$W$ + jets       & $2.9 \pm 0.4$ &  $1.1 \pm 0.2$ & $1.8 \pm 0.3$ & $<0.1$              & $<0.1$  \\
$t\bar{t}$         & $0.6 \pm 0.1$	&  $0.3 \pm 0.1$ & $0.3 \pm 0.1$ & $0.1 \pm 0.1$   & $<0.1$  \\ 
Diboson	       & $10.5 \pm 1.7$& $0.5 \pm 0.1$ & $8.5  \pm 1.4$& $0.4 \pm 0.1$   & $1.1\pm0.2$   \\
Multijet            & $<0.8$ & $<0.2$ & $<0.5$ & $<0.1$ & $<0.1$ \\
   \hline
   Background & & & & &\\
   Sum & $27.6 \pm 4.9$ & $7.5 \pm 1.2$ & $18.2 \pm 3.3$ & $0.6 \pm 0.1$  & $1.3 \pm 0.2$ \\
\hline
         Data       &    $22$  &  $5$   &  $15$ &  $0$ &  $2$     \\ 
 \hline\hline
\end{tabular}
\label{tab-events}
\end{center}
\end{table*}

As no significant excess in the data over the background is seen, limits were set on the $H^{++}$ production cross section using a modified frequentist approach ~\cite{collie}. The invariant mass of the two highest $p_T$ taus, $M(\tau_1,\tau_2)$, is used to discriminate signal from background.

\begin{figure}[ht]
\centering
\includegraphics[width=80mm]{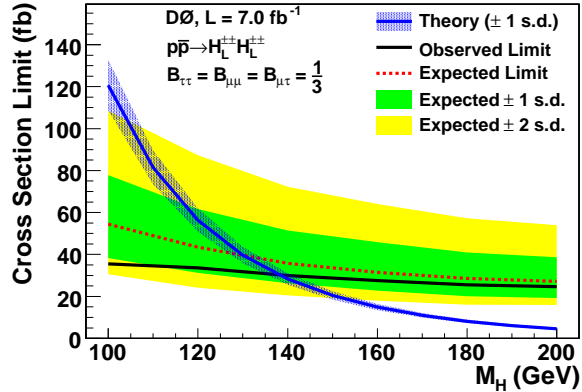}
\caption{The upper limit on the $H^{\pm\pm}_L H^{\pm\pm}_L$ pair production cross section. For the case where BR$(H^{\pm\pm} \rightarrow \tau\tau)$ =  BR$(H^{\pm\pm} \rightarrow \mu\mu)$ = BR$(H^{\pm\pm} \rightarrow \tau\mu)=1/3$. The black solid and dashed lines show the observed and expected limits respectively. The yellow and green regions are the one and two standard deviation bands and the blue band is the predicted cross section with associated uncertainty.  } \label{figure_six}
\end{figure}

The dominant sources of systematic uncertainties arise from, the uncertainty on the measured integrated luminosity, the uncertainties on the muon and tau identification including the uncertainty for applying the NN$_{\tau}$ and the uncertainties on the background cross sections for $Z/\gamma^*$, $W$+jets, $t\bar{t}$ and diboson production. There are also uncertainties from the trigger efficiency and on the signal acceptance due to the parton distribution function.  

Limits are set assuming several different values of the $H^{++}$ branching ratio (a) BR($H^{\pm\pm} \rightarrow \tau^{\pm}\tau^{\pm})$ = 1, (b) BR($H^{\pm\pm} \rightarrow \tau^{\pm}\mu^{\pm})$ = 1, (c) BR($H^{\pm\pm} \rightarrow \tau^{\pm}\tau^{\pm})$ = BR($H^{\pm\pm} \rightarrow \tau^{\pm}\mu^{\pm})$  = BR($H^{\pm\pm} \rightarrow \mu^{\pm}\mu^{\pm})$ = 1/3 and (d) assuming BR($H^{\pm\pm} \rightarrow \tau^{\pm}\tau^{\pm})$  + BR($H^{\pm\pm} \rightarrow \mu^{\pm}\mu^{\pm})$ =  1. For (d) the BR($H^{\pm\pm} \rightarrow \tau^{\pm}\tau^{\pm})$ is ranged in 10\% increments from 100\% BR to taus to 100\% BR to muons. In order to cover all this parameter space this analysis is combined with the a search from the D0 collaboration ~\cite{mmmm} for the pair production of $H^{++}$ decaying into 4$\mu$ final state, using  1.1 fb$^{-1}$ of integrated luminosity. The distribution of the invariant mass of the two highest $p_T$ muons, as determined by this analysis, including all systematics and their uncertainties and correlations here included in the limit setting procedure. For all the above scenarios, limits were set for both a left-handed and right-handed doubly charged Higgs apart from scenario (c) as the motivation is a model in which only left-handed Higgs exist.  Figure 5 shows the determined limits for scenario (c). Figure 6 shows the limits for scenario (d) for both a left-handed and right-handed doubly charged Higgs. 
The limits determined are summarized in Table II and are the most stringent limits on a pair produced doubly charged Higgs in these decay channels and the only limits set a hadron collider in  the case when it is assumed BR($H^{\pm\pm} \rightarrow \tau^{\pm}\tau^{\pm})$ = 1.  

\begin{figure*}[ht]
\centering
\includegraphics[width=80mm]{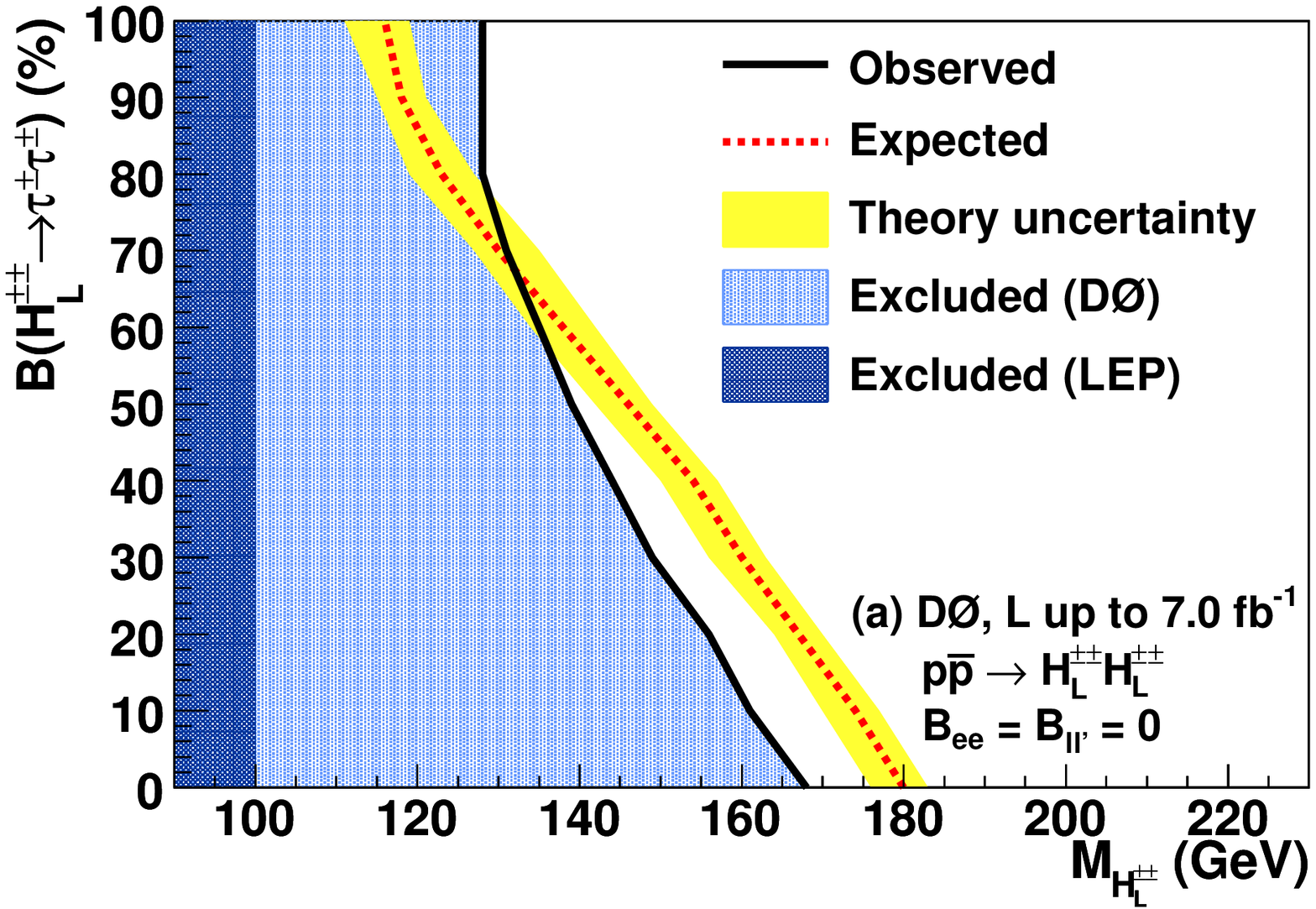}
\includegraphics[width=80mm]{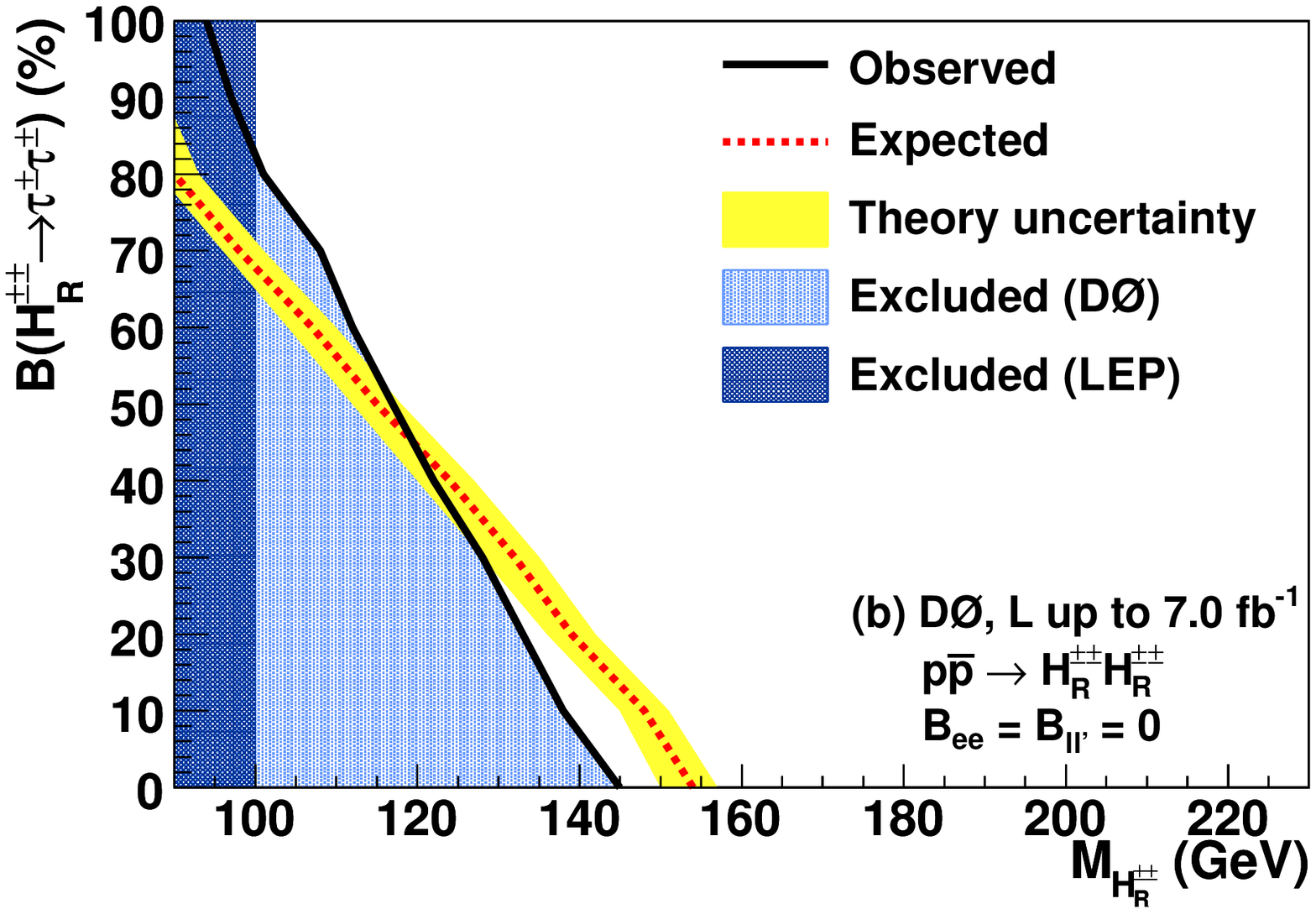}
\caption{The expected and observed exclusion regions at 95\% confidence level in the BR$(H^{\pm\pm} \rightarrow \tau^{\pm}\tau^{\pm}$), $M(H^{+++}$ plane, assuming  BR$(H^{\pm\pm} \rightarrow \tau^{\pm}\tau^{\pm}$) + BR$(H^{\pm\pm} \rightarrow \mu^{\pm}\mu^{\pm}) = 1$. For (a) a left-handed Higgs and (b) a right-handed Higgs. The yellow shows the uncertainty on the NLO calculation of the signal cross section.  } \label{figure_seven}
\end{figure*}

\begin{table}[htbp]
 \renewcommand{\tabcolsep}{0.5mm}
\caption{Expected and observed limits on
$M(\Hpm)$ (in GeV) 
for left and right-handed $\Hpm$ bosons.
Only left-handed states are considered for the model
that assumes equality of branching fractions into
$\tau\tau$, $\mu\tau$, and $\mu\mu$ final states. 
We only derive limits if the expected limit on $M(\Hpm)$ is $\ge 90$~GeV.
  }
 \begin{center}
   \begin{tabular}{c|cccc}
\hline\hline
   Decay & \multicolumn{2}{c}{$\HpmL$}& \multicolumn{2}{c}{$\HpmR$} \\
  \cline{2-5}
   & expected & observed & expected & observed \\
${\cal B}(\Hpm\to\tau^{\pm}\tau^{\pm})=1$ & $116$ & $128$ &   \\
${\cal B}(\Hpm\to\mu^{\pm}\tau^{\pm})=1$ & $149$ & $144$ & $119$ & $113$ \\
Equal $\BR$ into  &  &  & & \\
$\tau^{\pm}\tau^{\pm},\mu^{\pm}\mu^{\pm},\tau^{\pm}\mu^{\pm}$ & $130$ & $138$ & & \\
\hline 
${\cal B}(\Hpm\to\mu^{\pm}\mu^{\pm})=1$ & $180$ & $168$ & $154$ & $145$ \\
\hline\hline
 \end{tabular}
 \label{tab-limits}
 \end{center}
\end{table}

\begin{acknowledgments}
The author would like to thank the organizers of DPF as well as the collaborators at D0 and CDF who contributed to these analyses. 

\end{acknowledgments}

\bigskip 

\end{document}